\documentclass{article}

\usepackage{ifpdf}
\usepackage[latin1]{inputenc}
\usepackage[T1]{fontenc}
\usepackage{times}
\usepackage{graphicx}
\usepackage{url}
\usepackage{multirow}
\usepackage{amsmath,amssymb,amsfonts}
\usepackage{subfigure}
\usepackage[margin=1in]{geometry} 
\usepackage{tikz}

\newcommand {\br}[1]{\left(#1\right)}

\newcommand {\OMIT}[1]{}

\begin{document}

\title{Classification of arrayCGH data using a fused SVM}

\author{Franck Rapaport\\Institut Curie,Paris\\ Inserm,U900,Paris\\ and Ecole des Mines de Paris,Fontainebleau\\ \texttt{Franck.Rapaport@curie.fr} \and  \and Emmanuel Barillot\\Institut Curie,Paris\\ Inserm,U900,Paris\\ and Ecole des Mines de Paris,Fontainebleau\\ \texttt{Emmanuel.Barillot@curie.fr} \and Jean-Philippe Vert\\Institut Curie,Paris\\ Inserm,U900,Paris\\ and Ecole des Mines de Paris,Fontainebleau\\ \texttt{Jean-Philippe.Vert@ensmp.fr}}

\maketitle

\begin{abstract}

Motivation:
Array-based comparative genomic hybridization (arrayCGH) has recently become a popular tool to identify DNA copy number variations along the genome. These profiles are starting to be used as markers to improve prognosis or diagnosis of cancer, which implies that methods for automated supervised classification of arrayCGH data are needed. Like gene expression profiles, arrayCGH profiles are characterized by a large number of variables usually measured on a limited number of samples. However, arrayCGH profiles have a particular structure of correlations between variables, due to the spatial organization of BACs along the genome. This suggests that classical classification methods, often based on the selection of a small number of discriminative features, may not be the most accurate methods and may not produce easily interpretable prediction rules.

Results:
We propose a new method for supervised classification of arrayCGH data. The method is a variant of support vector machine (SVM) that incorporates the biological specificities of DNA copy number variations along the genome as prior knowledge. The resulting classifier is a sparse linear classifier based on a limited number of regions automatically selected on the chromosomes, leading to easy interpretation and identification of discriminative regions of the genome. We test this method on three classification problems for bladder and uveal cancer, involving both diagnosis and prognosis. We demonstrate that the introduction of the new prior on the classifier leads not only to more accurate predictions, but also to the identification of known and new regions of interest in the genome.

Availability:
All data and algorithms are publicly available.
\end{abstract}

\section{Introduction}

Genome integrity is essential to cell life and is ensured in normal cells by a series of checkpoints, which enable DNA repair or trigger cell death to avoid abnormal genome cells to appear. The p53 protein is probably the most prominent protein known to play this role. When these checkpoints are bypassed the genome may evolve and undergo alterations to a point where the cell can become premalignant and further genome alterations lead to invasive cancers.

This genome instability has been shown to be an enabling characteristic of cancer \cite{Hanahan2000}, and almost all cancers are associated with genome alterations. These alterations may be single mutations, translocations, or copy number variations (CNVs). A CNV can be a deletion or a gain of small or large DNA regions, an amplification, or an aneuploidy (change in chromosome number).

Many cancers present recurrent CNVs of the genome, like for example monoploidy of chromosome 3 in uveal melanoma \cite{Speicher1994}, loss of chromosome 9 and amplification of the region of cyclin D1 (11q13) in bladder carcinomas~\cite{Blaveri2005}, loss of 1p and gain of 17q in neuroblastoma~\cite{Bown2001,Roy2002}, EGFR amplification and deletion in 1p and 19q in gliomas \cite{Idbaih2007}, or amplifications of 1q, 8q24, 11q13, 17q21-q23, and 20q13 in breast cancer~\cite{Yao2006}. Moreover associations of specific alterations with clinical outcome have been described in many pathologies \cite{Lastowska1997}.

Recently array-based comparative genomic hybridization (arrayCGH) has been developed as a technique allowing rapid mapping of CNVs of a tumor sample at a genomic scale \cite{Pinkel1998}. The technique was first based on arrays using a few thousands of large insert clones (like BACs, and with a Mb range resolution) to interrogate the genome, and then improved with oligonucleotide based arrays consisting of several hundreds of thousands features, taking the resolution down to a few kb \cite{Gershon2005}. Many projects have since been launched to systematically detect genomic aberrations in cancer cells \cite{Beers2006,Chin2006,Shing2003}. 

The etiology of cancer and the advent of arrayCGH make it natural to envisage building classifiers for prognosis or diagnosis based on the genomic profiles of tumors.Building classifiers based on expression profiles is an active field of research, but little attention has been paid yet to genome-based classification\OMIT{, with the exception of~\cite{Chin2006}, \cite{Jones2004} and \cite{O'Hagan2003}}. \cite{Chin2006} select a small subset of genes and apply a k-nearest neighbor classifier to discriminate between estrogen-positive and estrogen-negative patients, between high-grade patients and low-grade patients and between bad prognosis and good prognosis for breast cancer. \cite{Jones2004} reduce the DNA copy number estimates to ``gains'' and ``losses'' at the chromosomal arm resolution, before using a nearest centroid method for classifying breast tumors according to their grade. As underlined in \cite{Chin2006}, the classification accuracy reported in \cite{Jones2004} is better than the one reported in \cite{Chin2006}, but still remains at a fairly high level with as much as $24\%$ of misclassified samples in the balanced problem. This may be related to the higher resolution of the arrays produced by \cite{Jones2004}. Moreover, the approach used by \cite{Jones2004} produces a classifier difficult to interpret as it is unable to detect any deletion or amplification that occur at the local level. \cite{O'Hagan2003} used a support vector machine (SVM) classifier \OMIT{and removed all the BACs where there was at least one missing value for at least one sample, excluding as much as 102 of 976 BACs from the analysis} using as variables all BAC ratios without any missing value. They were able to identify key CNAs.

The methods developed so far either ignore the particularities of arrayCGH and the inherent correlation structure of the data \cite{O'Hagan2003}, or drastically reduce the complexity of the data at the risk of filtering out useful information \cite{Jones2004,Chin2006}. In all cases, a reduction of the complexity of the data or a control of the complexity of the predictor estimated is needed to overcome the risk of overfitting the training data, given that the number of probes that form the profile is often several orders of magnitude larger than the number of samples available to train the classifier.
In this paper we propose a new method for supervised classification, specifically designed for the processing of arrayCGH profiles. In order not to miss potentially relevant information that may be lost if the profiles are first processed and reduced to a small number of homogeneous regions, we estimate directly a linear classifier at the level of individual probes. Yet, in order to control the risk of overfitting, we define a prior on the linear classifier to be estimated. This prior encodes the hypothesis that (i) many regions of the genome should not contribute to the classification rule (sparsity of the classifier), and (ii) probes that contribute to the classifier should be grouped in regions on the chromosomes, and be given the same weight within a region. This a priori information helps reducing the search space and produces a classification rule that is easier to interpret. This technique can be seen as an extension of SVM where the complexity of the classifier is controlled by a penalty function similar to the one used in the fused lasso method to enforce sparsity and similarity between successive features \cite{Tibshirani2005}. We therefore call the method a \textit{fused SVM}. It produces a linear classifier that is piecewise constant on the chromosomes, and only involves a small number of loci without any a priori regularisation of the data. From a biological point of view, it avoids the prior choice of recurrent regions of alterations, but produces \textit{a posteriori} a selection of discriminant regions which are then amenable to further investigations.

We test the fused SVM on several public datasets involving diagnosis and prognosis applications in bladder and uveal cancer, and compare it with a more classical method involving feature selection without prior information about the organization of probes on the genome. In a cross-validation setting, we show that the classification rules obtained with the fused SVM are systematically more accurate that the rules obtained with the classical method, and that they are also more easily interpretable.

\section{Methods}
In this section we present an algorithm for the supervised classification of arrayCGH data. This algorithm, which we call \emph{fused SVM}, is motivated by the linear ordering of the features along the genome and the high dependancy in behaviour of neighbouring features. The algorithm itself estimates a linear predictor by borrowing ideas from recent methods in regression, in particular the \emph{fused lasso} \cite{Tibshirani2005}. We start by a rapid description of the arrayCGH technology and data, before presenting the fused SVM in the context of regularized linear classification algorithms.

\subsection{ArrayCGH data}
ArrayCGH is a microarray-based technology that allows the quantification of the DNA copy number of a sample at many positions along the genome in a single experiment. The array contains thousands to millions of spots, each of them consisting of the amplified or synthesized DNA of a particular region of the genome. The array is hybridized with the DNA extracted from a sample of interest, and in most cases with (healthy) reference DNA. Both samples have first been labelled with two different fluorochromes, and the ratio of fluorescence of both fluorochromes is expected to reveal the ratio of DNA copy number at each position of the genome. The log-ratio profiles can then be used to detect the regions with abnormalities (log-ratio significantly different of 0), corresponding to gains (if the log-ratio is significantly superior to 0) or losses (if it is significantly inferior to 0).

The typical density of arrayCGH ranges from 2400 BAC features in the pioneering efforts, corresponding to one approximately 100 kb probe every Mb \cite{Pinkel1998}, up to millions today, corresponding to one 25 to 70bp oligonucleotide probe every few kb, or even tiling arrays \cite{Gershon2005}.

There are two principal ways to represent arrayCGH data: as a log-ratio collection, or as a collection of status (lost, normal or gained, usually represented as -1, 0 and 1 which correspond to the sign of the log ratio). The status representation has strong advantages over the log-ratio as it reduces the complexity of the data, provides the scientist with a direct identification of abormalities and allows the straightforwad detection of recurrent alterations. However, converting ratios into status is not always obvious and often implies a loss of information which can be detrimental to the study: for several reasons such as heterogeneity of the sample or contamination with healthy tissue (which both result in cells with different copy numbers in the sample), the status may be difficult to infer from the data, whereas the use of the ratio values avoids this problem. In particular, if we want to use arrayCGH for discriminating between two subtypes of tumors or between tumors with different future evolution, all tumors may share the same important genomic alterations that are easily captured by the status assignment and at the same differences between the types of tumors may be better predicted by more subtle signals. Therefore, we consider below an arrayCGH profile as a vector of log-ratios for all probes in the array.

\subsection{Classification of arrayCGH data}\label{sec:cgh}

While much effort has been devoted to the analysis of single arrayCGH profiles, or populations of arrayCGH profiles in order to detect genomic alterations shared by the samples in the population, we focus on the supervised classification of arrayCGH. The typical problem we want to solve is, given two populations of arrayCGH data corresponding to two populations of samples, to design a classifier that is able to predict which population any new sample belongs to. This paradigm can be applied for diagnosis or prognosis applications, where the populations are respectively samples of different tumor types, or with different evolution. Although we only focus here on binary classification, the techniques can be easily extended to problems involving more than two classes using, for example, a series of binary classifiers trained to discriminate each class against all others.

While accuracy is certainly the first quality we want the classifier to have in real diagnosis and prognosis application, it is also important to be able to interpret it and understand what the classification is based on. Therefore we focus on linear classifiers, which associate a weight to each probe and produce a rule that is based on a linear combination of the probe log-ratios. The weight of a probe roughly corresponding to its contribution in the final classification rule, and therefore provides evidence about its importance as a marker to discriminate the populations. In should be pointed out, however, that when correlated features are present, the weight of a feature is not directly related to the individual correlation of the feature with the correlation, hence some care should be taken for the interpretation of linear classifier.

In most applications of arrayCGH classification, it can be expected that only a limited number of regions on the genome should contribute to the classification, because most parts of the genome may not differ between populations. Moreover, the notion of discriminative regions suggest that a good classifier should detect these regions, and typically be piecewise constant over them. We show below how to introduce these prior hypotheses into the linear classification algorithm.

\subsection{Linear supervised classification}
Let us denote by $p$ the number of probes hybridized on the arrayCGH. The result of an arrayCGH competitive hybridization is then a vector of $p$ log-ratios, which we represent by a vector $x$ in the vector space $\mathcal{X} = \mathbb{R}^p$ of possible arrayCGH profiles. We assume that the samples to be hybridized can belong to two classes, which we represent by the labels $-1$ and $+1$. The classes typically correspond to the disease status or the prognosis of the samples. The aim of binary classification is to find a decision function that can predict the class $y \in \{-1,+1\}$ of a data sample $x \in \mathcal{X}$. Supervised classification uses a database of samples $x_1, ..., x_n \in \mathcal{X}$ for which the labels $y_1,..., y_n \in \{-1,+1\}$ are known in order to construct the prediction function. We focus on \emph{linear} decision functions, which are defined by functions of the form $f(x) =  w^{\top}x$ where $w^{\top}$ is the transpose of a vector $w\in\mathbb{R}^d$. The class prediction for a profile $x$ is then $+1$ if $f(x)\geq 0$, and $-1$ otherwise. Training a linear classifier amounts to estimating a vector $w\in\mathbb{R}^d$ from prior knowledge and the observation of the labeled training set.

The training set can be used to assess whether a candidate vector $w$ can correctly predict the labels on the training set; one may expect such a $w$ to correctly predict the classes of unlabeled samples as well. This induction principle, sometimes referred to as \emph{empirical risk minimization}, is however likely to fail in our situation where the dimension of the samples (the number of probes) is typically larger than the number of training points. In such a case, many vectors $w$ can indeed perfectly explain the labels of the training set, without capturing any biological information. These vectors are likely to poorly predict the classes of new samples. A well-known strategy to overcome this overfitting issue, in particular when the dimension of the data is large compared to the number of training points available, is to look for \emph{large-margin} classifiers constrained by \emph{regularization} \cite{Vapnik1998}. A large-margin classifier is a prediction function $f(x)$ that not only tends to produce the correct sign (positive for labels $+1$, negative for class $-1$), but also tends to produce large absolute values. This can be formalized by the notion of \emph{margin}, defined as $yf(x)$: large-margin classifiers try to predict the class of a sample with large margin. Note that the prediction is correct if the margin is positive. The margin can be thought of as a measure of confidence in the prediction given by the sign of $f$, so a large margin is synonymous with a large confidence. Training a large-margin classifier means estimating a function $f$ that takes large margin values on the training set. However, just like for the sign of $f$, if $p>n$ then it is possible to find vectors $w$ that lead to arbitrarily large margin on all points of the training set. In order to control this overfitting, large-margin classifiers try to maximize the margin of the classifier on the training set under some additional constraint on the classifier $f$, typically that $w$ is not too ``large''. In summary, large-margin classifiers find a trade-off between the objective to ensure large margin values on the training set, on the one hand, and that of controlling the complexity of the classifier, on the other hand. The balance in this trade-off is typically controlled by a parameter of the algorithm.

More formally, large-margin classifiers typically require the definition of two ingredients:
\begin{itemize}
\item A \emph{loss function} $l(t)$ that is ``small'' when $t\in\mathbb{R}$ is ``large''. From the loss function one can deduce the \emph{empirical risk} of a candidate vector $w$, given by the average loss function applied to the margins of $w$ on the training set:
\begin{equation}\label{eq:remp}
R_{emp}(w) = \frac{1}{n}\sum_{i=1}^n l(y_{i}w_{i}^\top x)\,.
\end{equation}
The smaller the empirical risk, the better $w$ fits the training set in the sense of having a large margin. Typical loss functions are the hinge loss $l(t) = \max(0,1-t)$ and the logit loss $l(t) = \log\br{1+e^{-t}}$.
\item A \emph{penalty function} $\Omega(w)$ that measures how ``large'' or how ``complex'' $w$ is. Typical penalty functions are the $L_{1}$ and $L_{2}$ norms of $w$, defined respectively by $||w||_{1} = \sum_{i=1}^p |w_{i}|$ and $||w||_{2} = \br{\sum_{i=1}^p w_{i}^2}^{\frac{1}{2}}$.
\end{itemize}
Given a loss function $l$ and a penalty function $\Omega$, large-margin classifiers can then be trained on a given training set by solving the following constrained optimization problem:
\begin{equation}\label{eq:optimisation}
\min_{w \in \mathbb{R}^p}R_{emp}(w) \textrm{ subject to }\Omega(w)\leq \mu\,,
\end{equation}
where $\mu$ is a parameter that controls the trade-off between fitting the data, i.e., minimizing $R_{emp}(f)$, and monitoring the regularity of the classifier, i.e., monitoring $\Omega(w)$. Examples of large-margin classifiers include the support vector machine (SVM) and kernel logistic regression (KLR) obtained by combining respectively the hinge and logit losses with the $L_{2}$ norm penalization function \cite{Cortes1995,Boser1992,Vapnik1998}, or the $1$-norm SVM when the hinge loss is combined with the $L_{1}$ loss . 

The final classifier depends on both the loss function and the penalty function. In particular, the penalty function is useful to include prior knowledge or intuition about the classifier one expects. For example, the $L_{1}$ penalty function is widely used because it tends to produce sparse vectors $w$, therefore performing an automatic selection of features. This property has been successfully used in the context of regression \cite{Tibshirani1996Regression}, signal representation \cite{Chen1998Atomic}, survival analysis \cite{Tibshirani1997}, logistic regression \cite{Genkin2007,Krishnapuram2004bayesian}, or multinomial logistic regression \cite{Krishnapuram2005Sparse}, where one expects to estimate a sparse vector.

\subsection{Fused lasso}

Some authors have proposed to design specific penalty functions as a means to encode specific prior informations about the expected form of the final classifier. In the context of regression applied to signal processing, when the data is a time series, \cite{Land1996} propose to encode the expected positive correlation between successive variables by choosing a regularisation term that forces successive variables of the classifier to have similar weights. More precisely, assuming that the variables $w_{1},w_{2},\ldots,w_{p}$ are sorted in a natural order where many pairs of successive values are expected to have the same weight, they propose the \emph{variable fusion} penalty function:
\begin{equation}\label{eq:fusion}
\Omega_{fusion}(w) = \sum_{i = 1}^{n-1}|w_i - w_{i+1}|\,.
\end{equation}
Plugging this penalty function in the general algorithm (\ref{eq:optimisation}) enforces a solution $w$ with many successive values equal to each others, that is, tends to produce a piecewise constant weight vector. In order to combine this interesting property with a requirement of sparseness of the solution, \cite{Tibshirani2005} proposed to combine the lasso penalty and the variable fusion penalty into a single optimization problem with two constraints, namely:
\begin{eqnarray}\label{eq:fusedlasso}
&&\min_{w \in \mathbb{R}^n}R_{emp}(w)\nonumber\\
& \textrm{ under the constraints }&\sum_{i = 1}^{n-1}|w_i - w_{i+1}|\leq \mu \nonumber\\
&& \|w\|_1 \leq \lambda\,,\nonumber\\
\end{eqnarray}
where $\lambda$ and $\mu$ are two parameters that control the relative trade-offs between fitting the training data (small $R_{emp}$), enforcing sparsity of the solution (small $\lambda$) and enforcing the solution to be piecewise constant (small $\mu$). When the empirical loss is the mean square error in regression, the resulting algorithm is called \emph{fused lasso}. This method was illustrated in \cite{Tibshirani2005} with examples taken from gene expression datasets and mass spectrometry. Later, \cite{Tibshirani2007} proposed a tweak of the fused lasso for the purpose of signal smoothing, and illustrated it for the problem of discretising noisy CGH profiles. 

\subsection{Fused SVM}
Remembering from Section \ref{sec:cgh} that for arrayCGH data classification one typically expects the ``true'' classifier to be sparse and piecewise constant along the genome, we propose to extend the fused lasso to the context of classification and adapt it to the chromosome structure for arrayCGH data classification. The extension of fused lasso from regression to large-margin classification is obtained simply by plugging the fused lasso penaly constraints into a large-margin empirical risk in (\ref{eq:fusedlasso}). In what follows we focus on the empirical risk (\ref{eq:remp}) obtained from the hinge loss, which leads to a simple implementation as a linear program (see Section \ref{sec:implementation} below). The extension to other convex loss functions, in particular the logit loss function, results in convex optimization problems with linear constraints that can be solved with general convex optimization solvers \cite{Boyd2004Convex}.

In the case of arrayCGH data, a minor modification to the variable fusion penalty (\ref{eq:fusion}) is necessary to take into account the structure of the genome in chromosomes. Indeed, two successive spots on the same chromosome are prone to be subject to the same amplification and are therefore likely to have similar weights on the classifier; however, this positive correlation is not expected across different chromosomes. Therefore we restrict the pairs of successive features appearing in the function constraint (\ref{eq:fusion}) to be consecutive probes on the same chromosome.

We call the resulting algorithm a \emph{fused SVM}, which can be formally written as the solution of the following problem:
\begin{eqnarray}\label{eq:fused svm}
&&\min_{w \in \mathbb{R}^p}\sum_{i=1}^n \textrm{max}(0,1-y_iw^{\top}x_i) \nonumber\\
&\textrm{ under the constraints }&\sum_{i \sim j}|w_i - w_j|\leq \mu \nonumber\\
&& \sum_{i=1} |w_i| \leq \lambda\,,\nonumber\\
\end{eqnarray}
where $i \sim j$ if $i$ and $j$ are the indices of succesive spots of the same chromosome. As with fused lasso, this optimisation problem tends to produce classifiers $w$ with similar weights for consecutive features, while maintaining its sparseness. This algorithm depends on two paramters, $\lambda$ and $\mu$, which are typically chosen via cross-validation on the training set. Decreasing $\lambda$ tends to increase the sparsity of $w$, while decreasing $\mu$ tends to enforce successive spots to have the same weight.

This classification algorithm can be applied to CGH profiles, taking the ratios as features. Due to the effect of both regularisation terms, we obtain a sparse classification function that attributes similar weights to successive spots.

\subsection{Implementation of the fused SVM}\label{sec:implementation}
Introducing slack variables, the problem described in (\ref{eq:fused svm}) is equivalent to the following linear program :
\begin{eqnarray}
\min_{w, \alpha, \beta, \gamma}\sum_{i=1}^n\alpha_i &\textrm{ under the following constraints :}&\nonumber\\
\forall i=1,...,n &\alpha_i \geq 0&\nonumber\\
\forall i=1,...,n &\alpha_i \geq 1 - w^{\top}x_iy_i&\nonumber\\
&\displaystyle\sum_{i=1}^n\beta_i\leq\lambda&\nonumber\\
\forall i=1,...,p &\beta_i \geq w_i&\nonumber\\
\forall i=1,...,p &\beta_i \geq -w_i&\nonumber\\
&\displaystyle\sum_{k=1}^{q}\gamma_k\leq\mu&\nonumber\\
\forall i,j \textrm{ such that } i \sim j  &\gamma_k \geq w_i-w_j&\nonumber\\
\forall i,j \textrm{ such that } i \sim j  &\gamma_k \geq w_j-w_i&\nonumber\\
\end{eqnarray}
In our experiments, we implemented and solved this problem using Matlab and the SeDuMi 1.1R3 optimisation toolbox \cite{Sturm1999}.

\section{Data}

We consider two publicly available arrayCGH datasets for cancer research, from which we deduce three problems of diagnosis and prognosis to test our method. 

The first dataset contains arrayCGH profiles of 57 bladder tumor samples \cite{Stransky2006}. Each profile gives the relative quantity of DNA for 2215 spots. We removed the probes corresponding to sexual chromosomes, because the sex mismatch between some patients and the reference used makes the computation of copy number less reliable, giving us a final list of 2143 spots. We considered two types of tumor classification: either by grade, with 12 tumors of grade 1 and 45 tumors of higher grades (2 or 3) or by stage, with 16 tumors of stage Ta and 32 tumors of stage T2+. In the case of stage classification, 9 tumors with intermediary stage T1 were excluded from the classification.

The second dataset contains arrayCGH profiles for 78 melanoma tumors that have been arrayed on 3750 spots \cite{Trolet2008}. As for the bladder cancer dataset, we excluded the sexual chromosomes from the analysis, resulting in a total of 3649 spots. 35 of these tumors lead to the development of liver metastases within 24 months, while 43 did not. We therefore consider the problem of predicting, from an arrayCGH profile, whether or not the tumor will metastasize within 24 months.

In both datasets, we replaced the missing spots log-ratios by 0.
In order to assess the performance of a classification method, we performed a cross-validation for each of the three classification problems, following a leave-one-out procedure for the bladder dataset and a 10-fold procedure for the melanoma dataset. We measure the number of misclassified samples for different values of parameters $\lambda$ and $\mu$.

\section{Results}
In this section, we present the results obtained with the fused SVM on the datasets described in the previous section. As a baseline method, we consider a $L_{1}$-SVM which minimizes the mean empirical hinge loss suject to a constraint on the $L_{1}$ norm of the classifier in (\ref{eq:optimisation}). The $L_{1}$-SVM performs automatic feature selection, and a regularization parameter $\lambda$ controls the amount of regularization. It has been shown to be competitive classification method for high-dimensional data, such as gene expression data \cite{Zhu20041norm}. In fact the $L_{1}$-SVM is a particular case of our fused SVM, when the $\mu$ parameter is chosen large enough to relax the variable fusion constraint (\ref{eq:fusion}), typically by taking $\mu>2\lambda$. Hence by varying $\mu$ from a large value to $0$, we can see the effect of the variable fusion penalty on the classical $L_{1}$-SVM.

\subsection{Bladder tumors}

\begin{figure}
\begin{center}
  \subfigure{
    \begin{tikzpicture}
      \input{include/stranskygrade}
    \end{tikzpicture}
  }
  \centerline{
    \includegraphics[width=7cm]{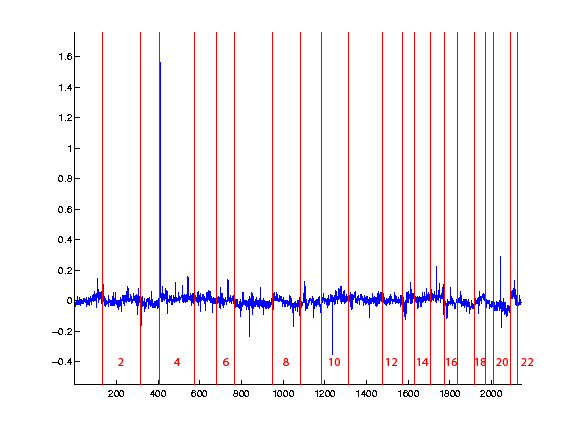}
    \includegraphics[width=7cm]{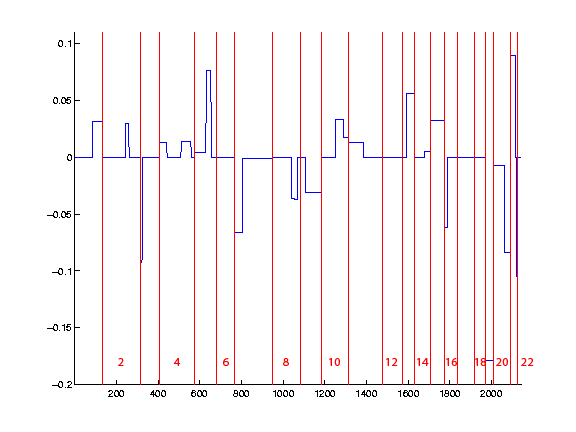}
  }
\end{center}
\caption[Bladder cancer dataset with grade classification]{The figure on the upper side represents the number of misclassified samples in a leave-one-out error loop on the bladder cancer dataset with the grade labelling, with its color scale for different values of the parameters $\lambda$ and $\mu$ which vary logarithmically along the axes. The weights of the best classifier, for classical $L_{1}$-SVM (left) and for fused SVM (right) are ordered and represented in a blue line, annotated with the chromosome separation (red line).}\label{fig:stranskygradeerror}
\end{figure}

\begin{figure}
\begin{center}
  \subfigure{
    \begin{tikzpicture}
      \input{include/stranskystage}
    \end{tikzpicture}
  }
  \centerline{
    \includegraphics[width=7cm]{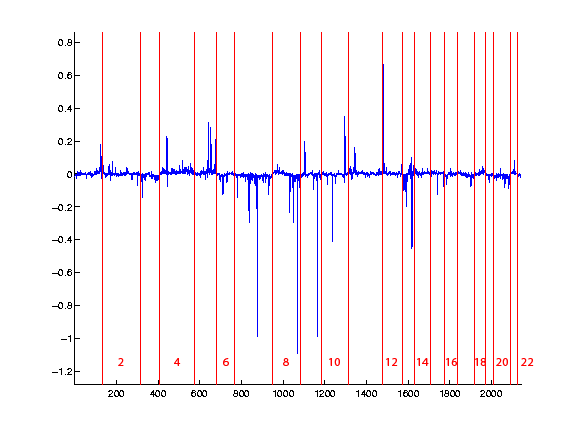}
    \includegraphics[width=7cm]{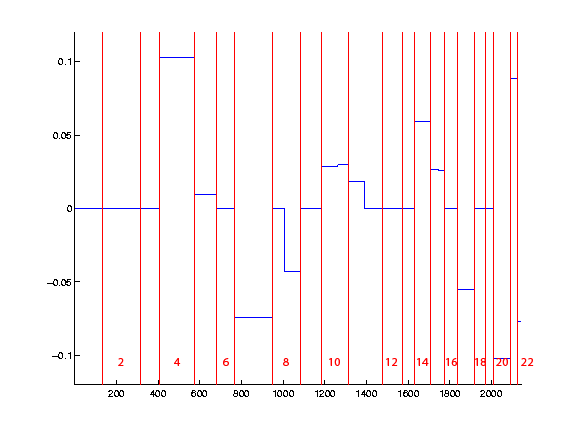}
  }
\end{center}
\caption[Bladder cancer dataset with stage classification]{The figure on the upper side represents the number of misclassified samples in a leave-one-out error loop on the bladder cancer dataset with the stage labelling, with its color scale, for different values of the parameters $\lambda$ and $\mu$ which vary logarithmically along the axes. The weights of the best classifier, for classical $L_{1}$-SVM (left) and for fused-SVM (right) are ordered and represented in a blue line, annotated with the chromosome separation (red line).}\label{fig:stranskystageerror}
\end{figure}

The upper plot of Figure \ref{fig:stranskygradeerror} show the estimated accuracy (by LOO) of the fused SVM as a function of the regularization parameters $\lambda$ and $\mu$, for the classification by grade of the bladder tumors. The lower left plot of Figure \ref{fig:stranskygradeerror} represents the best linear classifier found by the $L_{1}$-SVM (corresponding to $\lambda=256$), while the lower right plot shows the linear classifier estimated from all samples by the fused SVM when $\lambda$ and $\mu$ are set to values that minimise the LOO error, namely $\lambda=32$ and $\mu=1$. Similarly, Figure \ref{fig:stranskystageerror} shows the same results (LOO accuracy, $L_{1}$-SVM and fused SVM classifiers) for the classification of bladder tumors according to their stage.

In both cases, when $\mu$ is large enough to make the variable fusion inactive in (\ref{eq:fused svm}), then the classifier only finds a compromise between the empirical risk and the $L_1$ norm of the classifier. In other words, we recover the classical $L_1$ SVM with parameter $\lambda$. Graphically, the performance of the $L_1$ SVM for varying $\lambda$ can be seen on the upper side of each plot of the LOO accuracy in Figures  \ref{fig:stranskygradeerror} and \ref{fig:stranskystageerror}. Interestingly, in both cases we observe that the best performance obtained when both $\lambda$ and $\mu$ can be adjusted is much better than the best performance of the $L_{1}$ SVM, when only $\lambda$ can be adjusted. In the case of grade classification, the number of misclassified samples drops from $12$ (21\%) to $7$ (12\%), while in the case of stage classification it drops from $13$ (28\%) to $7$ (15\%). This suggests that the additional constraint that translates our prior knowlege about the structure of the spot positions on the genome is beneficial in terms of classifier accuracy.

As expected, there are also important differences in the visual aspects of the classifiers estimated by the $L_{1}$-SVM and the fused SVM. The fused SVM produces sparse and piecewise constant classifiers, amenable to further investigations, while it is more difficult to isolate from the $L_{1}$-SVM profiles the key features used in the classification, apart from a few strong peaks.

As we can see by looking at the shape of the fused SVM classifier in Figure \ref{fig:stranskygradeerror}, the grade classification function is characterised by non-null constant values over a few small chromosomal regions and numerous larger regions. Of these regions, a few are already known as being altered in bladder tumors, such as the gain on region 1q \cite{Corson2005}. Moreover some of them have already been shown to be correlated with grade, such as chromosome 7 \cite{Waldman1991}.

On the contrary, the stage classifier is characterised by only a few regions with most of them involving large portions of chromosomes. They concern mainly chromosome 4, 7, 8q, 11p, 14, 15, 17, 20, 21 and 22, with in particular a strong contribution from chromosomes 4, 7 and 20. These results on chromosomes 7, 8q, 11p and 20 are in good agreement with \cite{Blaveri2005} who identified the most common alterations according to tumor stage on a set of 98 bladder tumors.

\subsection{Melanoma tumors}
\begin{figure}
\begin{center}
  \subfigure{
    \begin{tikzpicture}
      \input{include/melanomaerror}
    \end{tikzpicture}
  }
  \centerline{
    \includegraphics[width=7cm]{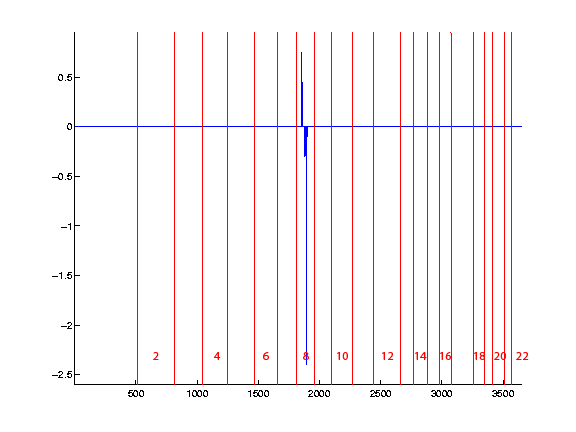}
    \includegraphics[width=7cm]{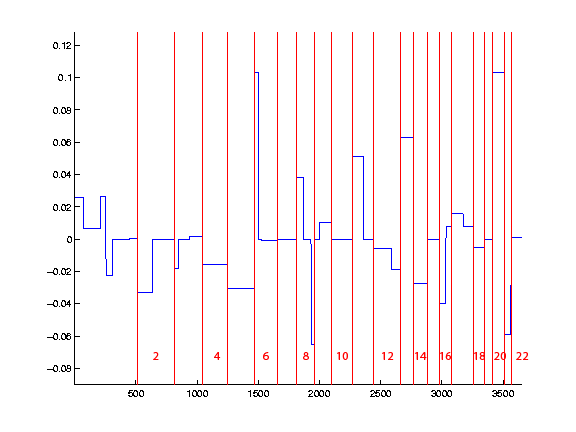}
  }
\end{center}
\caption[Uveal melanoma dataset]{The figure on the upper part represents the number of misclassified samples in a ten-fold error loop on the melanoma dataset. The weights of the best classifier, for classical $L_{1}$-SVM (left) and for fused SVM (right) are ordered and represented in a blue line, annotated with the chromosome separation (red line).}\label{fig:melanomaerror}
\end{figure}
Similarly to Figures \ref{fig:stranskygradeerror} and \ref{fig:stranskystageerror}, the three plots in Figure \ref{fig:melanomaerror} show respectively the the accuracy, estimated by 10-fold cross-validation, of the fused SVM as a function of the regularisation parameters $\lambda$ and $\mu$, the linear classifier estimated by the $L_{1}$-SVM when $\lambda$ is set to the value that minimizes the estimated error ($\lambda = 4$), and the linear classifier estimated by a fused SVM on all samples when $\lambda$ and $\mu$ are set to values that minimise the 10-fold error, namely $\lambda = 64$ and $\mu = 0.5$.

Similarly to the bladder study, the performance of the $L_{1}$-SVM without the fusion constraint can be retrieved by looking at the upper part of the plot of Figure \ref{fig:melanomaerror}. The fused classifier offers a slightly improved performance compared to the standard $L_{1}$-SVM (17 errors (22\%) versus 19 errors (24\%)), even though the amelioration seems more marginal compared to the improvement made with bladder tumors and the misclassification rate remains fairly high.

As for the bladder datasets, the $L_{1}$-SVM and fused SVM classifiers are markedly different. The $L_{1}$-SVM classifier is based only on a few BAC concentrated on chromosome 8, with positive weights on the 8p arm and negative weights on the 8p arm. These features are biologically relevant, and correspond to a known genomic alterations (loss of 8p and gain of 8q in metastatic tumors). The presence of a strong signal concentrated on chromosome 8 for the prediction of metastasis is in this case correctly captured by the sparse $L_{1}$-SVM, which explains its relatively good performance.

To the contrary, the fused SVM classifier is characterised by a many of CNAs, most of them involving large regions of chromosomes. Interestingly, we retrieve the regions whose alteration was already reported as recurrent events of uveal melanoma:  chromosomes 3, 1p, 6q, 8p, 8q, 16q. As expected the contributions of 8p and 8q are of opposite sign, in agreement with the common alterations of these regions: loss of 8p and gain of 8q in etastatic tumors. Interestingly the contribution of chromosome 3 is limited to a small region of 3p, and does not involve the whole chromosome as the frequency of chromosome 3 monosomy would have suggested. Note that this is consistent with works by \cite{Parrella2003} and \cite{Tschentscher2001} who delimited a small 3p regions from partial chromosome 3 deletion patients. On the other hand we also observe that large portions of other chromosomes have been assigned significant positive or negative weights, such as chromosomes 1p, 2p, 4, 5, 9q, 11p, 12q, 13, 14, 20, 21. To our knowledge, they do not correspond to previous observations, and may therefore provide interesting starting points for further investigations. 

\section{Discussion}

We have proposed a new method for the supervised classification of arrayCGH data. Thanks to the use of a particular regularization term that translates our prior assumptions into constraints on the classifier, we estimate a linear classifier that is based on a restricted number of spots, and gives as much as possible equal weights to spots located near each other on a chromosome. Results on real data sets show that this classification method is able to discriminate between the different classes with a better performance than classical techniques that do not take into account the specificities of arrayCGH data. Moreover, the learned classifier is piecewise constant and therefore lends itself particularly well to further interpretation, highlighting in particular selected chromosomal regions with particularly highly positive or negative weights.

From the methodological point of view, the use of regularized large-scale classifiers is nowadays widely spread, especially in the SVM form. Regularization is particularly important for ``small $n$ large $p$'' problems, i.e., when the number of samples is small compared to the number of dimensions. An alternative interpretation of such classifiers is that they correspond to maximum \emph{a posteriori} classifiers in a Bayesian framework, where the prior over classifier is encoded in our penalty function. It is not surprising, then, that encoding prior knowledge in the penalty function is a mathematically sound strategy that can be strongly beneficial in terms of classifier accuracy, in particular when few training samples are available. The accuracy improvements we observe on all classification datasets confirm this intuition. Besides the particular penalty function investigated in this paper, we believe our results support the general idea that engineering relevant priors for a particular problem can have important effects on the quality of the function estimated, and paves the way for further research on the engineering of such priors in combination with large-margin classifiers. As for the implementation, we solved a linear program for each values of regularization parameters $\lambda$ and $\mu$, but it would be interesting to generalize the recent works on path following algorithms to be able to follow the solution of the optimization problem when $\lambda$ and $\mu$ vary \cite{Efron2004Least}.

Another interesting direction of future research concerns the combination of heterogeneous data, in particular of arrayCGH and gene expression data. Gene expression variations contain indeed information complementary to CNV for the genetic aberrations of the dysfunctioning cell~\cite{Stransky2006}, and their combination is therefore likely to both improve the accuracy of the classification methods and shed new light on biological phenomena that are characteristic of each class. A possible strategy to combine such datasets would be to train a large-margin classifier with a particular regularization term that should be adequately designed.

\section*{Acknowledgement}
 We thank J\'er\^ome Couturier, Sophie Piperno-Neumann and Simon Saule, and the uveal melanoma group from Institut Curie. We are also grateful to Philippe Hup\'e for his help in preparing the data. This project was partly funded by the ACI IMPBIO Kernelchip and the EC contract ESBIC-D (LSHG-CT-2005-518192). FR and EB are members of the team ``Systems Biology of Cancer'', Equipe labellis\'ee par la Ligue Nationale Contre le Cancer.
\bibliographystyle{alpha}
\bibliography{biblio,../../../bibli/bibli}

\end{document}